\documentclass[prb,aps,twocolumn,a4paper,showpacs]{revtex4} 
\usepackage{epsfig}
\usepackage{umlaut}
\newcommand{\mmt}{Hg$_{\scriptstyle 1-x}$Mn$_{\scriptstyle x}$Te\,}
\newcommand{\cmta}{Hg$_{\scriptstyle 0.3}$Cd$_{\scriptstyle 0.7}$Te\,}

\newcommand{\czt}{Cd$_{\scriptstyle 0.96}$Zn$_{\scriptstyle 0.04}$Te\,}

\DeclareMathAlphabet{\mathitb}{OT1}{cmr}{bx}{sl}

\begin{document}
\title{Current heating of a magnetic 2DEG in 
Hg$_{1-x}$Mn$_x$Te/Hg$_{0.3}$Cd$_{0.7}$Te quantum wells}

\author{Y. S. Gui$^{1,2}$}
\author{C. R. Becker$^1$}
\email{becker@physik.uni-wuerzburg.de}
\author{J. Liu$^1$}
\author{M. K\"{o}nig$^1$}
\author{V.~Daumer$^1$}
\author{M. N. Kiselev$^1$}
\author{H. Buhmann$^1$}
\author{L. W. Molenkamp$^1$}
\affiliation{1; Physikalisches Institut der Universit\"at
W\"urzburg, Am  Hubland, 97074 W\"urzburg, Germany}
\affiliation{2; National Laboratory for Infrared Physics, 
Shanghai Institute of Technical
Physics,Chinese Academy of Sciences, Shanghai 200083, China}

\date{\today}

\begin{abstract}
Heating caused by electrons with excess kinetic energy
has been investigated in a magnetic two-dimensional
electron gas, M2DEG, in \mmt{}/\cmta{}(001) 
quantum wells. The
temperature of the Mn ions, $T_{\rm Mn}$, has been determined by the node 
positions in the beating pattern in Shubnikov-de Haas oscillations. 
The experimental dependence of  $T_{\rm Mn}$ on current and therefore
on electron temperature, is in excellent agreement with
a rate equation model. 
Results with this model show that the energy transfer rate from the
electrons to the Mn system is proportional to the Mn concentration.
 
\end{abstract}


\pacs{73.50.-h, 73.50.Gr, 73.61.Ga, 73.63.Hs}

\maketitle


Interesting new physics has resulted from the study of a two 
dimensional electron gas, 2DEG, which is subjected to
additional interactions. 
For example the Rashba spin-orbit, {\sc s-o},
interaction\cite{Rashba} has been the subject of numerous
investigations of III-V heterostructures.\cite{Das,Das90,Engels}
The Rashba {\sc s-o} coupling is particularily strong in narrow gap
HgTe quantum wells, QWs, with an inverted band structure.\cite{Zha01}
The recently introduced magnetic 2DEG\cite{Smorch97} in which magnetic ions
(usually Mn ions) are exchange coupled to the 2DEG is another example. 
Spin interactions,\cite{Crooker} and spin dependent transport and
localization,\cite{Smorch97} have been investigated in these systems.  
In addition, Rashba, Zeeman and Landau effects have been shown to
be of comparable magnitude in  a M2DEG found in Hg$_{1-x}$Mn$_x$Te
quantum wells.\cite{Gui2002}
The giant Zeeman splitting
caused by the $sp-d$ exchange interaction can be efficiently
suppressed by increasing the manganese temperature, while the
spin-orbit splitting only depends on the asymmetry of the
QW and is not sensitive to
temperature.\cite{Pfeffer1997}

Recently, Keller {\it et al.}\cite{Keller2001} have found an
efficient energy transfer from the photo-excited carriers to the Mn system,
which raised the temperature of the magnetic ion system in
a Zn$_{1-x}$Mn$_x$Se/Zn$_{1-y}$Be$_y$Se M2DEG. However, 
the power of the laser radiation is much
higher than that of the current normally employed 
in a magneto-transport experiment. Thus a comparison
of these two methods is of interest.
In this article, we report on the current heating of the 2DEG and the 
Mn ion system in Hg$_{1-x}$Mn$_x$Te/Hg$_{0.3}$Cd$_{0.7}$Te(001) QWs. 
Samples with different Mn concentrations have
been studied as a function of current by means of their Shubnikov-de
Haas, SdH, oscillations. It has been found that relatively small
current densities cause a strong suppression of the giant
Zeeman splitting of the conduction electrons, and this effect is
strongly dependent on the Mn content.


A series of $n$ type \mmt{}/\cmta{}(001) 
QWs were grown by molecular beam epitaxy, MBE, on
\czt{}(001) substrates. The QWs were 
modulation doped using CdI$_{2}$
as a doping material. The \mmt{} well width is 12~nm and
the \cmta{} barriers consist of a
5.5 nm thick spacer and a 9 nm thick doped layer. Standard Hall
bars with a width, $W$, of 200~$\mu$m were fabricated by 
wet chemical etching. A 200 nm thick
Al$_2$O$_{3}$ film was deposited on top of the structure, which
serves as an insulating layer. Ohmic indium contacts were fabricated
by thermal bonding. Magnetotransport measurements were carried
out using dc techniques with currents of 1
$\mu $A to 1.2~mA in magnetic fields ranging up to 7~T and bath
temperatures down to 1.4~K. The carrier concentrations and the
Hall mobilities were determined to be $3.3\times10^{12}$ cm$^{-2}$
and $5.2\times 10^{4}$ cm$^2$/Vs for Q1697 
($x = 0.015$), and $4.2\times10^{12}$ cm$^{-2}$ and $2.0\times 10^{4}$
cm$^2$/Vs for Q1715 ($x = 0.064$) at 4.2~K for zero gate voltage 
from low magnetic field Hall measurements.

\begin{figure}[t] 
  \centerline{{\epsfig{figure=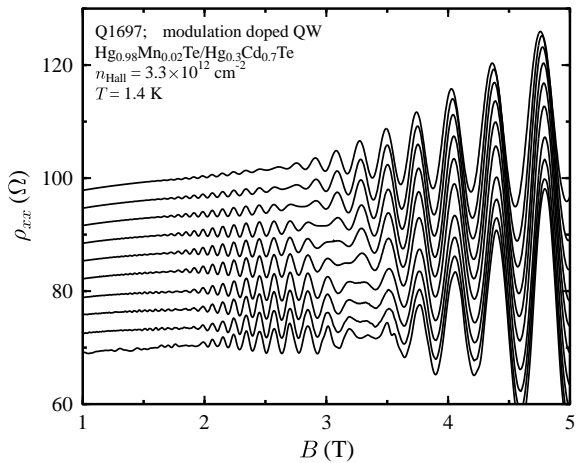,width=80mm}}} 
  \vspace{-4mm}
\caption{SdH oscillations at various currents 
for sample Q1697 ($x = 0.015$) at a lattice temperature, $T_{L}$,
of $1.4$~K. 
The curves from bottom to top
corresponded to the currents of 1, 10, 20, 50, 75, 100, 150, 200,
300 and 400 $\mu $A, respectively. The plots are
offset 2 $\Omega $ for clarity. } \label{q1697rxx-i}
\end{figure}
\begin{figure}[t] 
  \centerline{{\epsfig{figure=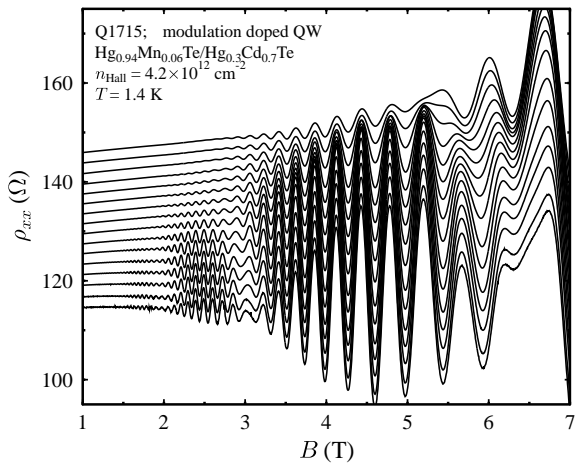,width=80mm}}} 
  \vspace{-4mm}
\caption{SdH oscillations at various currents 
for sample Q1715 ($x = 0.064$) at $T_{L} = 1.4$~K. 
The curves from bottom to top
corresponded to the currents of 5, 10, 20, 50, 75, 100, 150, 200,
300, 400, 500, 600, 800, 1000 and 1200 $\mu $A, respectively. The plots are
offset 2 $\Omega $ for clarity. } \label{q1715rxx-i}
\end{figure}

Fig.~\ref{q1697rxx-i} and \ref{q1715rxx-i} show a distinct beating pattern in
the SdH oscillations of samples Q1697 ($x = 0.015$) and Q1715 
($x = 0.064$) for various currents at 1.4~K. 
Nodes in the beating pattern shift with current but can no longer be resolved 
when the current exceeds 1 mA. These nodes correspond to 
the equivalence of the spin splitting energy and that of a
half integer multiple of the Landau splitting energy. In our
\mmt{}/\cmta{} samples, the total spin splitting energy is a combination
of Rashba\cite{Rashba} {\sc s-o} 
and giant Zeeman splitting energies.\cite{Gui2002}
The Rashba {\sc s-o} effect is due to the structure 
inversion asymmetry, SIA, of the quantum well which is small 
for the present samples. 
However the Rashba effect does not depend 
on temperature,\cite{Zha02} and consequently does not influence the results.
Only currents were employed which did not change the 2DEG
concentration and therefore the asymmetry of the QW, i.e., $\le
400~\mu$A and $\le 1.2~$mA for Q1697 and Q1715, respectively.  

\begin{figure}[t] 
  \centerline{{\epsfig{figure=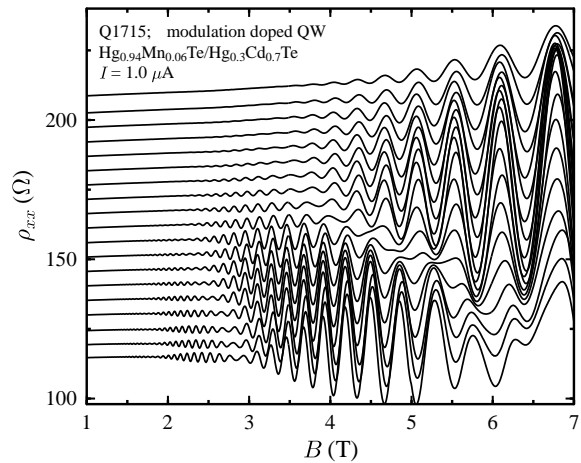,width=80mm}}} 
  \vspace{-4mm}
\caption{SdH oscillations for Q1715 ($x = 0.064$) 
at temperatures of 1.4, 2.0, 2.5, 3.0, 4.0, 5.0, 6.0, 7.0, 8.0, 10,
12, 14, 16, 18, 20, 23, 26, 31 and 41~K from bottom to top.
The plots are offset 5~$\Omega $ for clarity.} \label{q1715rxx-t}
\end{figure}

Only giant Zeeman splitting depends on the   
temperature of the Mn ions, $T_{Mn}$, according to the
phenomenological expression;\cite{Brandt1984,Smorchkova1997}
\begin{equation}
\label{eq1} E_{Z} = g_{0} \mu_{B}\,B - \frac{(\Delta E)_{\max }\,B_{5 /2}
\,(5g_{Mn} \mu{_B}B)}{(2k_B (T_{Mn} + T_0))}
\label{brill}
\end{equation}
where $g_{Mn} = 2$, $B_{5 / 2}(x)$ is
the Brillouin function for a spin of $S$=5/2, empirically
modified by using a rescaled temperature, $T_{Mn} + T_0$,
to account for antiferromagnetic spin-spin interaction,
and $(\Delta E)_{\max}$ is the saturated splitting energy caused
by $sp-d$ exchange interaction. If $(\Delta E)_{\max}$ and $T_0$
are known, then this modified Brillouin function can be used to 
determine the $T_{\rm Mn}$. To accomplish this the nodes in 
SdH oscillations at known lattice temperatures and a  
current of 1~$\mu$A were analysed, see for example Fig.~\ref{q1715rxx-t}. 
The resulting values of $(\Delta E)_{\max}$ and $T_0$ for Q1697 and
Q1715 together with other experimental and semi-empirical values for
$T_0$ from a magnetic susceptibility study of Hg$_{1-x}$Mn$_x$Te 
alloys,\cite{Bastard} are listed in Table~\ref{table1}. Obviously 
the experimental values of $T_0$ are in good agreement with those
of Bastard and Lewiner.\cite{Bastard}

\begin{table}[t]
\begin{minipage}{77mm}
\caption{Experimental values for $(\Delta E)_{max}$ and $T_0$ from
this work and experimental (a) and semi-empirical (b) values after
Bastard and Lewiner~\cite{Bastard}}
\begin{center}                                    
\begin{tabular}{c|c|c|c|c|c}\hline\hline
&$x$ & $(\Delta E)_{max}$ & $T_0$ & (a) $T_0^{exp}$ & (b) $T_0^{th}$\\ 
& & meV & K & K & K\\ \hline
Q1697 & 0.015 & $4.5 \pm 0.5$ & $2.5 \pm 0.5$ & 3.0 & 3.3 \\ 
Q1715 & 0.064 & $24 \pm 4$ & $5 \pm 1$ &  & 5.6 \\ \hline\hline 
\end{tabular}
\label{table1}
\end{center}
\end{minipage}
\end{table}

\begin{figure}[t] 
  \centerline{{\epsfig{figure=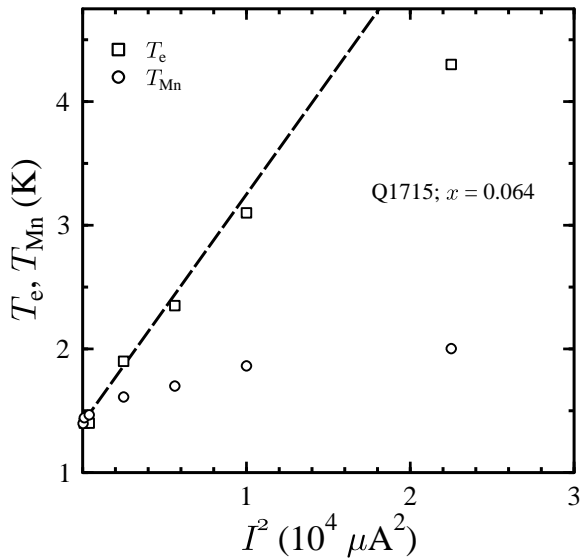,width=84mm}}} 
  \vspace{-4mm}
\caption{Temperature of the electrons and Mn ions in Q1715 ($x = 0.064$) 
versus current squared and electrical power. The straight dashed 
line represents the slope of $T_{e}(I^2)$ at low currents.}\label{q1715tetm-i2}
\end{figure}

\begin{figure}[h] 
  \centerline{{\epsfig{figure=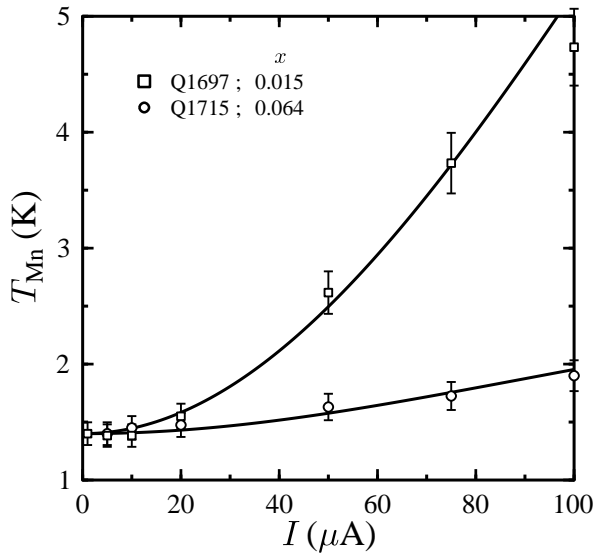,width=84mm}}} 
  \vspace{-4mm}
\caption{The Mn temperature versus current 
for Q1697 ($x = 0.015$) and Q1715 ($x = 0.064$) at 
$T_{L} = 1.4$~K and low current densities. The solid lines 
represent the calculated results of Eq.~\ref{tmn}} \label{tmn-i}
\end{figure}

Hence, $T_{\rm Mn}$ for the current heating experiment
 can be determined via Eq.~\ref{brill}.
Finally, the temperature of
the hot electrons has been determined from the temperature dependence
of the amplitudes of the fast Fourier transformation of the SdH 
oscillations, which scales as $X/sinh(X)$ where $X = f(m^*, B)\cdot T_e$.    
The results for $T_e$ and $T_{\rm Mn}$ are shown for Q1715 
in Fig.~\ref{q1715tetm-i2} as a function of current squared.  
The increase in electron temperature is proportional to $I^2$ at low
current values up to about $100~\mu$A. The energy relaxation time
associated with energy transfer to the 2DEG, $\tau_e$, should 
be related to their increase in temperature by the heat balance equation
\begin{equation}
c_{\rm v}\Delta T_e = (I/W)^2 \rho \tau_e
\label{heat}
\end{equation}
where the electronic heat capacity per unit area is given by 
$c_{\rm v} = (\pi^2/3)(k_{\rm B}T/E_{\rm F})n k_{\rm B}$, $W$ is
the width of the Hall bar, and $n$ the electron density. From the experimental 
values for low currents, which is indicated by the straight line 
in Fig.~\ref{q1715tetm-i2}, $n = 4.2 \times 10^{12}$~cm$^{-2}$ and 
$E_{\rm F} = 220$ meV, 
we find $\tau_e \sim 8 \times 10^{-11}$~s, which is a reasonable value
for a 2DEG in a QW.\cite{Molenkamp}

Deviation from the $I^2$ dependence at higher currents 
indicates that lattice heating
effects are no longer negligible; however, due 
to the large heat capacity of the HgTe lattice compared to that
of the 2DEG, any increase in the lattice temperature, $T_{\rm L}$, 
is assumed to be negligible, particularily for $I \le 100$~$\mu$A.

If the temperature difference between Mn and the lattice is small
then according to the rate equation model of 
K\"onig {\it et al.}\cite{Konig2000}
\begin{equation}
\frac{1}{T_{\rm Mn}} - \frac{1}{T_{\rm L}} = 
\frac{\tau_{\rm SL}}{\tau_{\rm SL} + (\tau_{\rm e-Mn} + \tau_{\rm s})}  
\bigg ( \frac{1}{T_{\rm e}} - \frac{1}{T_{\rm L}}\bigg )
\label{rate}
\end{equation} 
where  $\tau_{\rm SL}$,  
$\tau_{\rm e-Mn}$ and  $\tau_{\rm s}$ are the spin-lattice, 
electron-Mn ion and electron spin relaxation times, respectively.
By means of recursive substitution, this equation can be exactly rewritten as
\begin{equation}
\frac{T_{\rm Mn} - T_{\rm L}}{T_{\rm L}} = \frac{a y}{1 + y(1 - a)}
\label{tmn}
\end{equation} 
where 
\begin{equation}
a = \frac{\tau_{\rm SL}}{\tau_{\rm SL} + (\tau_{\rm e-Mn} + \tau_{\rm s})}  
\end{equation}
and 
\begin{equation}
y = \frac{T_{\rm e} - T_{\rm L}}{T_{\rm L}}
\end{equation} 
The $T_{\rm Mn} \approx T_{\rm L}$  condition is fulfilled at 
low current densities where in addition
$T_{\rm e} - T_{\rm L} = b I^2$, e.g., see Fig.~\ref{q1715tetm-i2}.
The subsequent application of Eq.~\ref{tmn} for these current
densities results in values which are in very 
good agreement with experiment as can be seen in Fig.~\ref{tmn-i}.
Furthermore, the ratio of $\tau_{\rm SL}/(\tau_{\rm e-Mn}+\tau_{\rm s})$ 
has been determined from the resulting value of $a$ and is listed in 
Table~\ref{table2} for both QWs. If $\tau_{\rm SL}$ and its Mn
dependence is assumed to be similar to that of other Mn containing II-VI
hererostructures\cite{Keller2001}, 
then the Mn dependence of the total characteristic time of 
energy transfer from the 2DEG to the Mn, $\tau$, 
can be determined according to
\begin{equation}
\frac{\tau_{\rm SL}^{0.015}}{\tau_{\rm SL}^{0.064}}\cdot
\frac{\tau^{0.064}}{\tau^{0.015}} = R_{\rm SL}\cdot R^{-1} = R_{\rm exp.}
\end{equation}
where $\tau = \tau_{\rm e-Mn} + \tau_{\rm s}$, and $R_{\rm SL}$,
$R^{-1}$ and $R_{\rm exp.}$ are the corresponding ratios. 
Using $R_{\rm exp.} \approx 12$ and values for other II-VI
materials,\cite{Keller2001} i.e.,  $R_{\rm SL} \approx 40$ to 100,
results in $R \approx 6$. 
In other words, the total relaxation time of the 2DEG is given by 
$\tau=(\tau_{\rm e-Mn}+\tau_{\rm s}) \propto 1/x$.

\begin{table}[t]
\begin{minipage}{77mm}
\caption{Experimental values for $a$, $b$ and 
$\tau_{\rm SL}/(\tau_{\rm e-Mn}+\tau_{\rm s})$, i.e., $a/(1-a)$.}
\begin{center}                                    
\begin{tabular}{c|c|c|c|c}\hline\hline
&$x$ &  $b(\mu$A$^{-2}$) & $a$ & $\tau_{\rm SL}/(\tau_{\rm e-Mn}+
\tau_{\rm s})$\\
\hline 
Q1697 & 0.015 & $5.2 \times 10^{4}$ & $0.94 \pm 0.02$ & $16.7 \pm 6.0$ \\ 
Q1715 & 0.064 & $1.9 \times 10^{4}$ & $0.58 \pm 0.10$ & \,\,$1.4 \pm 0.5$ \\ \hline\hline 
\end{tabular}
\label{table2}
\end{center}
\end{minipage}
\end{table}

At a given current, the temperature of the electrons is much 
higher than that of the Mn ions. 
As discussed in Refs.~\onlinecite{Keller2001} and
\onlinecite{Konig2000}, the hot carriers will lose some of their excess
energy to the Mn ion system via spin-flip scattering as well
as to the lattice. The heat loss from the Mn system to the lattice is
determined by the SLR. In very dilute
systems with $x < 0.01$, where Mn ions are isolated entities, the
spin-lattice relaxation time is extremely long. However, it
decreases by several orders of magnitude with an increasing
concentration of Mn ions, when clusters of of three or more magnetic ions are
formed.\cite{Scalbert1996} Under the influence of steady-state
heating, the resulting spin temperature, $T_{\rm Mn}$, will exceed the
lattice temperature. The temperature difference is
determined by the energy flux and the SLR time.


By analyzing nodes in the beating pattern of SdH oscillations
in a M2DEG in Hg$_{1-x}$Mn$_x$Te(001) QWs, the temperature 
of the electrons as well as that of the Mn ions have been determined.
When $T_{\rm Mn} - T_{\rm L}$ is small, experimental values of 
$T_{\rm Mn}$ are in excellent agreement with the predictions of a 
rate equation model. 
This leads to values for the ratio of $\tau_{\rm SL}/(\tau_{\rm e-Mn}+
\tau_{\rm s})$. 
These ratios are consistent with the expected shorter spin-lattice 
relaxation times at higher Mn concentrations. An analysis based on
this model results in the Mn
dependence of the total time for energy transfer from the 2DEG to the 
Mn ion system, which is inversely proportional to the Mn concentration. 


The authors thank D. R. Yakovlev for numerous
discussions. The financial support of the Deutsche
Forsch\-ungsgemeinschaft for SFB410, the DARPA SpinS program, the
National Natural Science Foundation of China, no. 10374094 (Y.S.G.),   
and the Chinese special funds for major state basic research,  
project no. G001CB3095 (Y.S.G.) is gratefully acknowledged.

\vspace{-5mm}



\begin{thebibliography}{99}
\bibitem{Rashba}E.I. Rashba, Fiz. Tverd. Tela {\bf 2}, 1224 (1960) [ Sov.
Phys. Solid State {\bf 2}, 1109 (1960)]; Yu. A. Bychkov and E.I
Rashba, Pis'ma Zh. Eksp. Teor. Fiz. {\bf 39}, 66 (1984) [JETP
Lett. {\bf 39}, 78 (1984)].

\bibitem{Das}B. Das, D. C. Miller, S. Datta, R. Reifenberger,
W. P. Hong, P. K. Bhattacharya, J. Singh, and M. Jaffe,
Phys. Rev. B {\bf 39}, 1411 (1989).
\bibitem{Das90}B. Das, S. Datta, and R. Reifenberger, Phys. Rev. B
{\bf 41}, 8278 (1990).

\bibitem{Engels}G. Engels,
J. Lange, Th. Sch\"{a}pers, and H. L\"{u}th,
Phys. Rev. B {\bf 55}, R1958 (1997);
J. Nitta, 
T. Akazaki, H. Takayanagi and T. Enoki, 
Phys, Rev. Lett. {\bf 78},1335 (1997);
C. M. Hu,
J. Nitta, T. Akazaki, H. Takayanagi, 
J. Osaka, P. Pfeffer and W. Zawadzki,
Phys. Rev. B {\bf 60}, 7736 (1999);
D. Grundler, Phys. Rev. Lett. {\bf 84}, 6074 (2000).

\bibitem{Zha01}X. C. Zhang, A. Pfeuffer-Jeschke,
K. Ortner, V. Hock, H. Buhmann, C. R. Becker, and G. Landwehr,
Phys. Rev. B {\bf 63} (24), 245305 (2001).\label{Zha01}


\bibitem{Smorch97}I. P. Smorchkova, N. Samarth, J. M. Kikkawa,
and D. D. Awschalom, Phys. Rev. Lett. {\bf 78}, 3571 (1997).

\bibitem{Crooker}S. A. Crooker, D. A. Tulchinsky, J. Levy,
D. D. Awschalom, R. Garcia and  N. Samarth,
Phys. Rev. Lett. {\bf 75}, 505 (1995).

\bibitem{Gui2002} Y. S. Gui, C. R. Becker, J. Liu, V. Daumer,
V. Hock, H. Buhmann and L. W. Molenkamp, Europhys. Lett. in press.

\bibitem{Pfeffer1997} P. Pfeffer, Phys. Rev. \textbf{B} 55, R7359 (1997);
P. Pfeffer and W. Zawadzki, Phys. Rev. \textbf{B} 59, R5312
(1999).

\bibitem{Keller2001} D. Keller, D. R. Yakovlev, B. K\"{o}nig, W. Ossau,
Th. Gruber, A Waag,  L. W. Molenkamp, and A.V. Scherbakov, 
Phys. Rev. \textbf{B} 65,
035313 (2001)

\bibitem{Zha02}X. C. Zhang, A. Pfeuffer-Jeschke,
K. Ortner, C. R. Becker, and G. Landwehr,
Phys. Rev. B {\bf 65} (24), 045324 (2002).\label{Zha02}


\bibitem{Brandt1984} N. B. Brandt and V. V. Moshchalkov, Advances
in Physics 33, 193(1984); J. K. Furdyna, J. Appl. Phys. 64, R29
(1988);

\bibitem{Smorchkova1997} I. P. Smorchkova, N. Samarth, J. M. Kikkawa
 and D. D. Awschalom, Phys. Rev. Lett. 78, 3571 (1997); I. P. Smorchkova,
 J. M. Kikkawa, N. Samarth and D. D.Awschalom, Physica \textbf{E} 2,
 381 (1998).

\bibitem{Bastard}G. Bastard nad C. Lewiner, J. Phys. C; Solid
St. Phys., {\bf 13}, 1469 (1980).


\bibitem{Molenkamp}H. van Houten, L. W. Molenkamp, C. W. J. Beenakker,
and C. T. Foxon, Semicond. Sci. Technol. {\bf 7}, B215 (1992).

\bibitem{Konig2000} B. K\"{o}nig, I. A. Merkulov, D. R. Yakovlev,
 W. Ossau, S. M. Ryabchenko, M. Kutrowski, T. Wojtowicz, G. Karczewski,
 and J. Kossut, Phys. Rev. \textbf{B} 61,16870 (2000)

\bibitem{Scalbert1996}D. Scalbert. phys. status solidi b 193, 189 (1996)

\end{thebibliography}
\end{document}